# OPTIMAL BLACK HOLES ARE THE COSMOLOGICAL OBJECTS, WHICH MINIMIZE VOLUME OF INFORMATION IN AREAS OF THE UNIVERSE AND IN THE UNIVERSE AS A WHOLE

Igor Gurevich

The Institute of Informatics Problems of the Russian Academy of Sciences,
Hetnet Consulting Corp., Moscow, Russia, iggurevich@gmail.com

**Abstract**

Black hole is called optimal if information content is minimal at the University region, consisting of usual substance and one($n$) black hole(s). Optimal black hole mass does not depend on the mass of the Universe region. Optimal black holes can exist when at least the two types of substance are available in the Universe: with non-linear and linear correspondence between information content and mass. Information content of optimal black hole is proportional to squared coefficient correlating information content with mass in usual substance and in inverse proportion to coefficient correlating information content with black hole mass. Concentration of mass in optimal black hole minimizes information content in the system "usual substance – black holes". Minimal information content of the Universe consisting of optimal black holes only is twice as less as information content available of the Universe of the same mass filled with usual substance only. Under the radiation temperature $T \approx 1E+12$ K the mass of optimal black holes that emerged in the systems "radiation – black hole" is equal to the mass of optimal black holes that emerged in the systems "hydrogen (protons) – black hole".

## 1. Introduction

By using informatics approach one can theoretically demonstrate the existence of black holes that minimize information content of arbitrary region of the Universe and place lower bound limitations on information content of the Universe. The origin and cause of optimal black holes existence is the occurrence of substance of two different types: with square-law and linear-law dependence of information content on mass. In the presence of substance of only one type, the optimal black holes do not exist. Following tasks are solved. The direct task: to discover an information minimum in system "usual substance-black a hole" at the given mass of usual substance and a black hole. The dual task: to discover a mass maximum in system "usual substance-black a hole" at the given information in usual substance and a black hole.

The present work reviews optimal black hole characteristics in the systems "radiation (photons) – black holes", "hydrogen (protons) – black holes" and in the system "several types of usual substance – black holes".

## 2. Definition of optimal black holes

Let us estimate the information content of the Universe region of mass $M_{Un\ rgn}$ under the arbitrary square-law relation between information and energy (mass) of the black hole $I_{bh} = \alpha \cdot M_{bh}^2$ and arbitrary linear-law relation between information and energy (mass) of usual substance $I_{us} = \beta \cdot M_{us}$ (under arbitrary non-negative coefficients $\alpha, \beta$) [1-3]. Aggregated information content of the Universe region of mass $M_{Un\ rgn}$, consisting of usual substance and one black hole is equal to

$$I_{Un\ rgn} = I_{bh} + I_{us} = \alpha \cdot M_{bh}^2 + \beta \cdot M_{us} =$$
$$= \alpha \cdot M_{bh}^2 + \beta \cdot (M_{Un\ rgn} - M_{bh}).$$

Let us find a condition for the minimum [4-8]:

$$\frac{\partial I_{Un\ rgn}}{\partial M_{bh}} = \frac{\partial (I_{bh} + I_{us})}{\partial M_{bh}} = \frac{\partial (\alpha \cdot M_{bh}^2 + \beta \cdot (M_{Un\ rgn} - M_{bh}))}{\partial M_{bh}} = 2\alpha \cdot M_{bh} - \beta = 0.$$

**Assertion 1.** $M_{Opt\ bh} = \frac{\beta}{2\alpha}$ is black hole mass, under which information content of the Universe region of mass $M_{Un\ rgn}$ consisting of usual substance and one black hole is minimal.

Let's call the black hole *optimal* under which information content is minimal at the University region of mass $M_{Un\ rgn}$, consisting of usual substance and one black hole. Black hole mass does not depend on the mass of the Universe region. Optimal black hole mass is proportional to coefficient correlating information content with usual substance mass and in inverse proportion to coefficient correlating information content with black hole mass.

*Note 1.* Let us estimate the information content of the Universe region of mass $M_{Un\ rgn}$ under the arbitrary square-law relation between information and energy (mass) of $n$ black holes $I_{bh\ n} = n\alpha \cdot M_{bh}^2$ and arbitrary linear-law relation between information and energy (mass) of usual substance $I_{us} = \beta \cdot M_{us}$ (under arbitrary non-negative coefficients $\alpha, \beta$) [1-3]. Aggregated information content of the Universe region of mass $M_{Un\ rgn}$, consisting of usual substance and $n$ black holes is equal to

$$I_{Un\ rgn} = nI_{bh} + I_{us} = n\alpha \cdot M_{bh}^2 + \beta \cdot M_{us} =$$
$$= n\alpha \cdot M_{bh}^2 + \beta \cdot (M_{Un\ rgn} - nM_{bh}).$$

Let us find a condition for the minimum:

$$\frac{\partial I_{Un\ rgn}}{\partial M_{bh}} = \frac{\partial(nI_{bh} + I_{us})}{\partial M_{bh}} = = \frac{\partial(n\alpha \cdot M_{bh}^2 + \beta \cdot (M_{Un\ rgn} - nM_{bh}))}{\partial M_{bh}} = 2n\alpha \cdot M_{bh} - n\beta = 2\alpha \cdot M_{bh} - \beta = 0.$$

$M_{Opt\ bh} = \frac{\beta}{2\alpha}$ is black hole mass, under which information content of the Universe region of mass $M_{Un\ rgn}$ consisting of usual substance and n black hole is minimal.

*Note 2.* Optimal black holes can exist when at least the two types of substance are available in the Universe: with non-linear (for instance, $I = \gamma \cdot M^\delta$ under $\gamma > 0, \delta > 1$) and linear correspondence between information content and mass.

**Assertion 2.** *Information content of optimal black hole is proportional to squared coefficient correlating information content with mass in usual substance and in inverse proportion to coefficient correlating information content with black hole mass:*

$$I_{Opt\ bh} = \frac{\beta^2}{4\alpha}\ bits.$$

*Note 3.* We will observe *a dual task* of definition of the maximum mass $M_{Un\ rgn}$ of system «usual substance - a black hole» at the given information content $I_{Un\ rgn}$ of the Universe region. Let's size up mass $M_{Un\ rgn}$ in the Universe region containing the given information content $M_{Un\ rgn}$ at the arbitrary square link between the information and energy (mass) of black hole $I_{bh} = \alpha \cdot M_{bh}^2$ or $M_{bh}^2 = \frac{1}{\alpha} I_{bh}$ ($M_{bh} = \frac{1}{\sqrt{\alpha}} \sqrt{I_{bh}}$), and the arbitrary linear link between the information and energy (mass) of usual substance $I_{us} = \beta \cdot M_{us}$ or $M_{us} = \frac{1}{\beta} I_{us}$ (at the arbitrary nonnegative coefficients $\alpha, \beta$). The information of the Universe region consisting of usual substance and one black hole is equal to $I_{Un\ rgn} = I_{bh} + I_{us}$. The mass $M_{Un\ rgn}$ of the Universe region consisting of usual substance and one black hole, is equal to

$$M_{Un\ rgn} = M_{bh} + M_{us} = \frac{1}{\sqrt{\alpha}} \sqrt{I_{bh}} + \frac{1}{\beta} I_{us} = = \frac{1}{\sqrt{\alpha}} \sqrt{I_{bh}} + \frac{1}{\beta}(I_{Un\ rgn} - I_{bh}).$$

We will discover an optimality requirement:

$$\frac{\partial M_{Un\ rgn}}{\partial I_{bh}} = \frac{\partial(M_{bh} + M_{us})}{\partial I_{bh}} =$$

$$= \frac{\partial(\frac{1}{\sqrt{\alpha}}\sqrt{I_{bh}} + \frac{1}{\beta}\cdot(I_{Un\ rgn} - I_{bh}))}{\partial I_{bh}} = \frac{1}{\sqrt{\alpha}}\cdot\frac{1}{2\sqrt{I_{bh}}} - \frac{1}{\beta} = 0.$$

As $\frac{\partial^2 M_{Un\ rgn}}{\partial I_{bh}^2} = -\frac{1}{\sqrt{\alpha}}\cdot\frac{1}{4I_{bh}^{3/2}} < 0$ the given requirement is a maximum requirement.

Further we have: $\frac{1}{\sqrt{\alpha}}\cdot\frac{1}{2\sqrt{I_{bh}}} = \frac{1}{\beta}$, $\frac{\beta}{2\sqrt{\alpha}} = \sqrt{I_{bh}}$ and the information content in a black hole of optimum mass is equal $I_{bh} = \frac{\beta^2}{4\alpha}$. The mass of a black hole at which mass $M_{Un\ rgn}$ of region the Universe is maximum at the given information content of the Universe region $I_{Un\ rgn}$, is equal to

$$M_{Opt\ bh} = \frac{1}{\sqrt{\alpha}}\sqrt{I_{bh}} = \frac{1}{\sqrt{\alpha}}\sqrt{\frac{\beta^2}{4\alpha}} = \frac{\beta}{2\alpha}.$$

**Assertion 1.a.** *The Information content in a black hole at which mass of the Universe region $M_{Un\ rgn}$ consisting of usual substance and one black hole and containing the given information content $I_{Un\ rgn}$, it is proportional to a square of coefficient linking an information content with mass in usual substance and inversely proportional to coefficient linking an information content with mass in a black hole:*

$$I_{Opt\ bh} = \frac{\beta^2}{4\alpha}\ bits.$$

**Assertion 2.a.** *The mass of a black hole at which the mass of field of Universe $M_{Un\ rgn}$ containing the given information content $I_{Un\ rgn}$ is maximum, consisting of usual substance and one black hole, is equal*

$$M_{Opt\ bh} = \frac{\beta}{2\alpha}.$$

Information contents and the masses gained at the solution of a direct task (a minimisation of information content in system «usual substance - a black hole» at the given mass of system - assertions 1, 2) and a dual task (a maximisation of mass of system «usual substance - a black hole» at the given information content) assertions 1.a, 2.a), coincide. Thereby concept of an optimum black hole is identical and all subsequent assertions and relationships also are identical.

As far as the black hole mass under which information content is minimal at the Universe region consisting of black hole and common substance does not depend neither on aggregated mass of the Universe (the Universe), nor on usual substance mass in the region under study, then the minimal information content of the Universe region (the Universe) can be reached if the Universe region (the Universe) consists of optimal black holes only. The maximum number of optimal black holes of the Universe region (the Universe) is equal to

$$N_{Opt\ Un\ rgn} = \frac{M_{Un\ rgn}}{M_{Opt\ bh}} = M_{Un\ rgn} \frac{2\alpha}{\beta} \quad (N_{Opt\ bh\ Un} = \frac{M_{Un}}{M_{Opt\ bh}} = M_{Un} \frac{2\alpha}{\beta}).$$

The minimal information level of the Universe region (the Universe) consisting of black holes only is equal to

$$I_{Un\ bh} = N_{Opt\ bh} \cdot I_{Opt\ bh} = M_{Un} \cdot \frac{2\alpha}{\beta} \frac{\beta^2}{4\alpha} = \frac{M_{Un} \cdot \beta}{2}.$$

*Note 4.* Hereafter for brevity sake we'll speak about the Universe, though the assertions and expressions are also true for arbitrary regions of the Universe.

**Assertion 3.** *Minimal information content of the Universe consisting of optimal black holes only is twice as less as information content available of the Universe of the same mass filled with usual substance only:*

$$I_{Un\ bh} = \frac{M_{Un} \cdot \beta}{2} \ bits.$$

Assertions 1-3 are true for any kind of linear dependence of information volume on usual substance mass. The important agents of usual substance are radiation and hydrogen. Let's consider next the optimal black holes characteristics in the systems "radiation (photons) – black holes", "hydrogen (protons) – black holes", "several types of usual substance – black holes".

## 3. The Universe filled with radiation

Let's consider the Universe filled with usual substance (radiation). The energy required for transfer, retrieval, record of one bit under the temperature $T$ can not be less than the value $E_{min} = kT\ ln2$ [9-10]. In accordance with the Einstein equation, the mass required for transfer, retrieval, record of one bit under the temperature $T$ can not be less than the value $M_{min} = (kT\ ln2)/c^2$. It is easily seen that for record of *1* bit under $T = 1$ K the mass no less than $M_{bit} = E_{bit}/c^2 \approx 10^{-23}$ *joule*/$(9 \cdot 10^{16}\ m^2/s^2) = 10^{-40}$ *kg* is needed.

As far as $\alpha = \frac{2\pi \cdot G}{\hbar \cdot c \cdot \ln 2}$, and for radiation $\beta = \frac{c^2}{k \cdot T \cdot \ln 2}$, then $M_{Opt\ bh} = \frac{\beta}{2\alpha} = \frac{\hbar \cdot c^3}{4\pi \cdot G \cdot k \cdot T}$ and information content of optimal black hole formed in the system "black hole + radiation" is equal to

$I_{Opt\ bh} = \frac{\beta^2}{4\alpha} = \frac{\hbar \cdot c^5}{8\pi \cdot G \cdot k^2 \cdot T^2 \cdot \ln 2}$ bit. Let's express the obtained result as the following assertion.

**Assertion 4.** *Concentration of mass* $M = \dfrac{\hbar \cdot c^3}{4\pi \cdot G \cdot k \cdot T}$ *of optimal black hole minimizes information content in the system "photons — black holes".*

We note that the mass of optimal black hole that emerged in the system "radiation – black hole" is in inverse proportion to radiation temperature.

The total number of optimal black holes $N_{Opt\,bh}$, in the Universe of mass $M_{Un}$ consisting of radiation and black holes is equal to

$$N_{Opt\,bh} = M_{Un} \frac{2\alpha}{\beta} = M_{Un} \frac{4\pi \cdot G \cdot k \cdot T}{\hbar \cdot c^3}.$$

**Assertion 5.** *The minimal possible information content of the Universe of mass $M_{Un}$ of the Universe of mass $M_{Un}$ consisting of radiation and black holes is equal* $M_{Un} \dfrac{c^2}{2 \cdot k \cdot T \cdot \ln 2}$ *to*

$$I_{Un\,min} = I_{Opt\,bh} \cdot N_{Opt\,bh} = \frac{M_{Un} \cdot \beta}{2} \text{ bits.}$$

Thus, the minimal possible information content of the Universe of the Universe of mass $M_{Un}$ consisting of radiation and black holes is proportional to the mass of the Universe, the speed of light squared, inversely proportional to Boltsman constant and the temperature of the Universe.

**Assertion 6.** *The Universe of mass $M_{Un}$, consisting of radiation and black holes, containing* $N_{Opt\,bh} = M_{Un} \dfrac{4\pi \cdot G \cdot k \cdot T}{\hbar \cdot c^3}$ *black holes of mass* $M_{Opt\,bh} = \dfrac{\hbar \cdot c^3}{4\pi \cdot G \cdot k \cdot T}$, *while only each of the black holes of the given mass contains the minimal possible information content equal to*

$$I_{Un\,min} = M_{Un} \frac{c^2}{2 \cdot k \cdot T \cdot \ln 2} \text{ bits.}$$

The minimal possible information content of the Universe of mass $M_{Un}$, consisting of radiation and black holes is proportional to the energy of the Universe, inversely proportional to Boltsman constant and the temperature of the Universe. It is notable that the minimal information content in the Universe does not depend neither on gravitation constant nor Plank constant

$$M_{Opt\,bh} = \frac{\hbar \cdot c^3}{4\pi \cdot G \cdot k \cdot T} = 9,09 \cdot 10^{25} \text{ g} = = 9,09 \cdot 10^{22} \text{ kg.}$$

The mass of black hole under which the information minimum is gained in the Universe of mass $M_{Un}$ consisting of radiation and black holes does not depend on the gross mass of the Universe and is equal to $9,09 \cdot 10^{22}$ kg. It is approximately one seventieth of the Earth mass which is equal to $6 \cdot 10^{24}$ kg. Information content of optimal black hole is equal to $I_{Opt\,bh} \approx 1,26 \cdot 10^{62}$ bits. Our

Universe can contain about $10^{29}$ black holes. The minimal information content in the Universe of the mass equal to the mass $\approx 10^{52}$ kg of our Universe, consisting of $10^{29}$ optimal black holes, and only of these, s equal to

$$I_{Un\ min} = N_{Opt\ bh} \cdot I_{Opt\ bh} = \frac{c^2}{2 \cdot k \cdot T \cdot \ln 2} \approx 1{,}56 \cdot 10^{91} \text{ bits.}$$

At $T \approx 2{,}7^0 10^n K$ the mass of an optimum black hole is approximately equal $9{,}09 \cdot 10^{22-n}$ kg, the information volume in an optimum black hole is approximately equal to $10^{62-n}$ bits. So at $T \approx 2{,}7^0 \cdot 10^{10} K$ (the nucleosynthesis beginning) mass of an optimum black hole is approximately equal $10^{13}$ kg, the information volume in an optimum black hole is approximately equal to $10^{52}$ bits.

**Assertion 7.** *Information content of the Universe of mass $M_{Un} = 10^{52}$ kg consisting of radiation and black holes ranges within $10^{91} \leq I_{Un\ M} \leq 10^{120}$ bits.*

**Assertion 8.** *Information content available in the Universe of mass $M_{Un}$, consisting of radiation and black holes ranges within*

$$M_{Un} \frac{c^2}{2 \cdot k \cdot T \cdot \ln 2} \leq I_{Un} \leq M_{Un}^2 \frac{2\pi \cdot G}{\hbar \cdot c \cdot \ln 2}.$$

## 4. The Universe filled with hydrogen (protons)

Let us consider the Universe filled with usual substance (hydrogen). Наполненную обычным веществом (водородом). Wave function of proton with upward-directed spin [11]

$$\varphi\left(P, s_z = \frac{1}{2}\right) = \frac{1}{\sqrt{18}}(2|p\uparrow\rangle|n\downarrow\rangle|p\uparrow\rangle + 2|p\uparrow\rangle|p\uparrow\rangle|n\downarrow\rangle + 2|n\downarrow\rangle|p\uparrow\rangle|p\uparrow\rangle -$$
$$- |p\uparrow\rangle|p\downarrow\rangle|n\uparrow\rangle - |p\uparrow\rangle|n\uparrow\rangle|p\downarrow\rangle - |p\downarrow\rangle|n\uparrow\rangle|p\uparrow\rangle - |n\uparrow\rangle|p\downarrow\rangle|p\uparrow\rangle - |n\uparrow\rangle|p\uparrow\rangle|p\downarrow\rangle - |p\downarrow\rangle|p\uparrow\rangle|n\uparrow\rangle).$$

Uncertainty (information content) of proton structure is equal to 2,837 bit. Having in mind the uncertainty of spin orientation it is necessary to add 1 bit - 3,837 bit. Information content in quarks (1 bit in each) – 3 bit. Colour information content - 2,585 bit. The total uncertainty (information content) of proton is contained in the proton structure, qurks and colour and equal to 9,422 bit. Hydrogen atom in the ground state ($|IV\rangle = \frac{|+-\rangle + |-+\rangle}{2}$ [12]) contains 11,422 bit (1 bit in atomic structure, 9,422 bit in proton and 1 bit in electron).

As far as $\alpha = \dfrac{2\pi \cdot G}{\hbar \cdot c \cdot \ln 2}$, and for hydrogen atoms $\beta = \dfrac{11{,}422}{m_в} \approx \dfrac{11{,}422}{m_p}$, then

$$M_{Opt\,bh} = \dfrac{\beta}{2\alpha} = \dfrac{11{,}422 \cdot \ln 2 \cdot \hbar \cdot c}{4\pi \cdot m_p \cdot G}$$ and information content of optimal black hole formed in the system "black hole + hydrogen" is equal to

$$I_{Opt\,bh} = \dfrac{\beta^2}{4\alpha} = \dfrac{(11{,}422)^2 \cdot \hbar \cdot c \cdot \ln 2}{8\pi \cdot m_p^2 \cdot G} \text{ bits.}$$

Let us define the obtained result as the following assertion.

**Assertion 9.** *Concentration of mass* $M_{Opt\,bh} = \dfrac{11{,}422 \cdot \ln 2 \cdot \hbar \cdot c}{4\pi \cdot m_p \cdot G}$ *in the optimal black hole minimizes information content in the system "hydrogen – black holes".*

The total number of optimal black holes $N_{Opt\,bh}$, in the Universe of mass $M_{Un}$ consisting of hydrogen atoms and blck holes is equal to

$$N_{Opt\,bh} = M_{Un} \dfrac{2\alpha}{\beta} = \dfrac{4\pi \cdot M_{Bc} \cdot m_p \cdot G}{11{,}422 \cdot \ln 2 \cdot \hbar \cdot c}.$$

**Assertion 10.** *The minimal possible information content of the Universe of mass $M_{Un}$, consisting of hydrogen atoms and black holes is equal to*

$$I_{Un\,\min} = I_{Opt\,bh} \cdot N_{Opt\,bh} = \dfrac{M_{Un} \cdot \beta}{2} \; M_{Un} \dfrac{11{,}422}{2 \cdot m_p} = 5{,}7 \dfrac{M_{Un}}{m_p} \text{ bits.}$$

Thus, the minimal possible information content of the Universe, of the Universe of mass $M_{Un}$ consisting of hydrogen atoms and black holes is proportional to the Universe mass and inversely proportional to two masses of hydrogen atom (proton).

**Assertion 11.** *The Universe of mass $M_{Un}$, consisting of hydrogen atoms and black holes, containing* $N_{Opt\,bh} = M_{Un} \dfrac{4\pi \cdot M_{Un} \cdot m_p \cdot G}{11{,}422 \cdot \ln 2 \cdot \hbar \cdot c}$ *black holes of mass*

$M_{Opt\,bh} = \dfrac{11{,}422 \cdot \ln 2 \cdot \hbar \cdot c}{4\pi \cdot m_p \cdot G}$ and only each of the black holes of the given mass contains the minimal possible information content equal to

$$I_{Un\,min} = 5{,}7 \dfrac{M_{Un}}{m_p} \text{ bits.}$$

The minimal possible information content of the Universe, of the Universe of mass $M_{Un}$, consisting of hydrogen atoms and black holes is proportional to the mass of the Universe, inversely proportional to hydrogen mass (proton). It is important to note that the minimal information content of the Universe consisting of hydrogen atoms and black holes does not depend on gravity constant, the speed of light and Plank constant.
Let us evaluate the mass of optimal black hole

$$M_{Opt\,bh} = \dfrac{11{,}422 \cdot \ln 2 \cdot \hbar \cdot c}{4\pi \cdot m_p \cdot G} = 1{,}78 \cdot 10^{14} \text{ г} = 1{,}78 \cdot 10^{11} \text{ kg.}$$

Black hole mass, under which the minimum information is gained at the Universe, the Universe of mass $M_{Un}$ consisting of hydrogen atoms and black holes, does not depend on the gross mass of the Universe and is equal to $1{,}78 \cdot 10^{11}$ kg. This is about the boundary mass of primary black hole equal to $\approx 10^{12}$ kg. Information content of optimal black hole is equal to

$$I_{Opt\,bh} = \dfrac{\beta^2}{4\alpha} = \dfrac{(11{,}422)^2 \cdot \hbar \cdot c \cdot \ln 2}{8\pi \cdot m_p^2 \cdot G} = 3{,}76 \cdot 10^{38} \text{ bits.}$$

There can be about $10^{41}$ optimal black holes in our Universe. The minimal information content of the Universe whose mass is equal to mass of our Universe of $\approx 10^{52}$ kg, consisting of $10^{41}$ optimal black holes and only of these, is equal to

$$I_{Un\,min} = 5{,}7 \dfrac{M_{Un}}{m_p} \approx 3{,}31 \cdot 10^{79} \text{ bits.}$$

**Assertion 12.** *Information content of the Universe of mass $M_{Un} = 10^{52}$ kg, consisting of hydrogen atoms and black holes lies in the range $10^{79} \leq I_{Un\,M} \leq 10^{120}$ bits.*

**Assertion 13.** *Information content of the Universe of mass $M_{Un}$, consisting of hydrogen atoms and black holes lies in the range*

$$5,7 \frac{M_{Un}}{m_p} \leq I_{Un} \leq M_{Un}^2 \cdot \frac{2\pi \cdot G}{\hbar \cdot c \cdot \ln 2} \text{ bits.}$$

## 5. Maximim information content of the Universe

The maximum possible information content is available at the Universe if the latter looks like one black hole of mass $M_{Un}$ [13]: $I_{Un\,max} = \beta \cdot M_{Un} = \frac{2\pi \cdot G}{\hbar \cdot c \cdot \ln 2} M_{Un}^2$ bits. The maximum information content of the Universe is proportional to squared mass of the Universe, gravity constant, inversely proportional to Plank constant, the speed of light and does not depend on Boltsman constant and the temperature of the Universe. The maximum information content of the Universe represented by one black hole whose mass is equal to mass of our Universe ($\approx 10^{52}$ kg), is equal to $I_{Un\,max} \approx 10^{120}$ bits.

## 6. Comparison of the characteristics of optimal black holes in the systems "radiation (photons) – black holes", "hydrogen (protons) – black holes"

Let us compare the characteristics of optimal black holes in the systems "radiation (photons) – black holes", "hydrogen – black holes".

The following table presents information about the characteristics of optimal black holes in the systems "black holes – radiation", "black holes – hydrogen".

| Characteristic | System "radiation (photons) – black holes" | System "hydrogen (protons) – black holes" | Note |
|---|---|---|---|
| Factor of proportionality "information – squared mass" for black holes | $\alpha = \dfrac{2\pi G}{\hbar c \ln 2}$ | $\alpha = \dfrac{2\pi G}{\hbar c \ln 2}$ | |
| Factor of proportionality "information–mass" for usual substance | $\beta = \dfrac{c^2}{kT \ln 2}$ | $\beta = \dfrac{11,422}{m_{_{\text{в}}}} \approx \dfrac{11,422}{m_p}$ | |
| Expression for optimal black hole mass | $M_{Opt\,bh} = \dfrac{\hbar c^3}{4\pi GkT}$ | $M_{Opt\,bh} = \dfrac{11,422\, \hbar c \ln 2}{4\pi m_p G}$ | In the system "radiation – black hole", the optimal black hole mass is |
| | | | inversely proportional to radiation temperature. |
| Estimtion of optimal black hole mass | $9{,}09 \cdot 10^{22}$ kg | $1{,}78 \cdot 10^{11}$ kg | In the system "hydrogen – black hole" the optional black hole mass is $\approx 10^{12}$ times less than the optimal black hole mass in the system "radiation – black hole". |
| Expression for estimation of information content of optimal black hole | $I_{Opt\ bh} = \dfrac{\hbar c^5}{8\pi G k^2 T^2 \ln 2}$ | $I_{Opt\ bh} = \dfrac{(11{,}422)^2 \hbar c \ln 2}{8\pi m_p^2 G}$ | |
| Estimation of information content of optimal black hole | $I_{Opt\ bh} \approx 1{,}26 \cdot 10^{62}$ bits | $I_{Opt\ bh} \approx 3{,}76 \cdot 10^{38}$ bits | In the system "hydrogen – black hole", information content of the optimal black hole is $\approx 10^{24}$ times less than information volume of the optimal black hole in the system "radiation – black hole". |
| Expression for estimation of the number of optimal black holes in the system with given mass | $N_{Opt\ bh} = \dfrac{4\pi M_{Un} G k T}{\hbar c^3}$ | $N_{Opt\ bh} = \dfrac{4\pi M_{Bc} m_p G}{11{,}422\ \hbar c \ln 2}$ | |
| Estimation of the number of optimal black holes in the Universe | $N_{Opt\ bh} \approx 1{,}1 \cdot 10^{29}$ | $N_{Opt\ bh} \approx 10^{41}$ | The number of optimal black holes in the Universe that emerged in the systems "hydrogen – black hole" is $\approx 10^{12}$ times larger than the number of optimal black holes that emerged in the systems "radiation – black hole". |

| Expression for estimation of the minimum information content in the system with given mass | $M_{Un} \cdot \dfrac{c^2}{2 \cdot k \cdot T \cdot \ln 2}$ | $5{,}7 \dfrac{M_{Un}}{m_p}$ | |
|---|---|---|---|
| Estimation of the minimum information content of the Universe | $I_{Un\,\min} \approx 1{,}7 \cdot 10^{91}$ | $I_{Un\,\min} \approx 3{,}31 \cdot 10^{79}$ | The minimum information content of the Universe consisting of black holes that emerged in the systems "hydrogen – black hole" is $\approx 10^{12}$ times less than the minimum information content of the Universe consisting of black holes that emerged in the systems "radiation – black hole". |
| Expression for estimation of the maximum information content in the system with a given mass | $M_{Un}^2 \dfrac{2\pi G}{\hbar c \ln 2}$ | $M_{Un}^2 \dfrac{2\pi G}{\hbar c \ln 2}$ | |
| Estimation of the maximum information content of the Universe | $I_{Un\,\max} \approx 1{,}9 \cdot 10^{120}$ | $I_{Un\,\max} \approx 1{,}9 \cdot 10^{120}$ | |

*Note 5.* Under the radiation temperature $T = m_p \dfrac{c^2}{k \cdot \ln 2 \cdot 9{,}422} = 1{,}555 \cdot E+12$ K the mass of optimal black holes that emerged in the systems "radiation – black hole" is equal to the mass of optimal black holes that emerged in the systems "hydrogen (protons) – black hole". By virtue of the fact that stable hydrogen atoms do not exist under high temperatures, then in such case the calculations have been done with respect to protons.

*Note 6.* In the period of transition from the Universe with predominant radiation to the Universe with predominant substance [14] ($10^4 > T > 10^3$), the mass of optimal black hole in the system "radiation – black hole" changes from 2,45E+19kg to 2,45E+20kg.

## 7. The systems consisting of black holes and several types of usual substance

Let us consider the systems consisting of black holes and several types of usual substance, for instance, of various kinds of particles.

From informatics point of view, various types of usual substance differ in coefficient $\beta_i$ standing for the use of mass per 1 bit of information $I_i = \beta_i M$. $\beta_i$ - denotes information content of the given type of usual substance.

**Assertion 14.** *The mass of optimal black hole in the system "sereval type of usual substnce – black holes" under which information content of the system under consideration is getting minimized, is defined by the minimal factor of proportionality $\beta_{i0} = \min_i \beta_i$. The optimal black hole corresponds to the system "usual substance of type $i0$ - black holes".*

**Assertion 15.** *The mass of black hole under which information content is minimal, meaning information under which information content is minimal in the system "several type of usual substance –black holes", is equal to*

$$M_{Opt\ bh} = \frac{\beta_{i0}}{2\alpha}.$$

**Assertion 16.** *Information content of optimal black hole in the system "several types of usual substance- black holes" is proportional to squared minimal coefficient correlating information content with mass in different types of usual substance and inversely proportional to coefficient correlating information content with mass in black hole:*

$$I_{Opt\ bh} = \frac{\beta_{i0}^2}{4\alpha}\ bits.$$

As the black hole mass under which information content is minimal in *the system "several types of usual substance – black holes"*, does not depend on neither on the total mass of the system $M$, $M$, nor on the mass of usual substance, then the minimum information content of the Universe is gained if the system consists of optimal black holes only. The maximum number of optimal black holes in *the system "several types of usual substance – black holes"*

$$N_{Opt\ bh} = \frac{M}{M_{Opt\ bh}} = M\frac{2\alpha}{\beta_{i0}}.$$

The minimum information content in *the system "several types of usual substance – black holes"*, consisting of black holes only is equal to

$$I_M = N_{Opt\ bh} \cdot I_{opt\ bh} = M \cdot \frac{2\alpha}{\beta}\frac{\beta^2}{4\alpha} = \frac{M \cdot \beta_{i0}}{2}.$$

**Assertion 17.** *If optimum black holes are formed of various types of atoms of usual substance or a mix of various types of atoms of usual substance masses of optimum black holes and volumes of the information in them are approximately identical.*

8. **Conclusion**

8.1. The occurrence of substance of two types: with square-law and linear-law dependence of information content on mass - is the origin and cause of optimal black holes existence that minimize information content in the arbitrary region of the Universe as well as of the Universe as a whole.

8.2. Optimal black hole mass is proportional to coefficient correlating information content with black hole mass. Optimal black hole mass does not depend on the mass of the region of the Universe where it is being formed.

8.3. The present work considers the characteristics of optimal black holes in the systems "radiation (photons) – black holes", "hydrogen (protons – black holes" and in the systems "several types of usual substance – black holes".

8.4. Because the black hole mass under which information content is minimal at the Universe region consisting of black hole and usual substance, does not depend neither on the gross mass of the Universe region (Universe), nor on the mass of usual substance in the region under study, then the minimum information content at the Universe region (Universe) is gained if the Universe region (Universe) consists of optimal black holes only.

8.5. Black hole mass under which the minimum information content is gained at the Universe of mass $M_{Un}$ consisting of radiation and black holes, does not depend on the gross mass of the Universe and is equal to $9{,}09 \cdot 10^{22}$ kg. It is approximately one seventieth of the Earth mass. Information content of optimal black hole is equal to $\approx 1{,}26 \cdot 10^{62}$ bits. Our Universe can contain about $10^{29}$ such optimal black holes. The minimum information content of the Universe is equal to $\approx 10^{91}$ bits.

8.6. Black hole mass under which the minimum of information content is gained at the Universe consisting of hydrogen atoms and black holes, does not depend on the gross mass of the Universe and is equal to $1{,}78 \cdot 10^{11}$ kg. Information content of optimal black hole is equal to $3{,}76 \cdot 10^{38}$ bits. The minimum information content of the Universe whose mass is equal to the mass of our Universe that is $\approx 10^{52}$ kg, consisting of approximately $10^{41}$ optimal black holes and only of these, is equal to $\approx 10^{79}$ bits.

Optimal black hole mass in the systems "several types of usual substance – black holes" under which information content in the system under consideration is minimized, is determined by the type of usual substance with minimal factor of proportionality $\beta_{i0} = \min_{i} \beta_i$. If optimum black holes are formed of various types of atoms of usual substance or a mix of various types of atoms of usual substance masses of optimum black holes and volumes of the information in them are approximately identical.

8.7. The maximum possible information content is available at the Universe if the latter appears to be one black hole. In this case it is equal to $\approx 10^{120}$ bits.

**Acknowledgements.**

The author thanks N.Kardashev, I.Novikov, I.Sokolov, S.Shorgin, K.Kolin, V.Sinitsin, V.Lipunov, L.Gindilis, M.Abubekerov, and especially A.Panov for the shown interest and support of this direction, and also for useful discussions of stated ideas.

**References**

1. *Penrose R.* The Emperor's New Mind. Oxford University Press. 1989.
2. *Shapiro S.L., Teucolsky S.A.* Black holes, white dwarfs, and neutron stars. The pfysics of compact objects. Cornell University, Itaca. New York. A Wiley-Intersciense pablication. John Wiley & Sons. 1983.
3. *Novicov I.D., Frolov V.P.* The pfysics of black holes. (In Russian). Science. Moscow. 1986. 328 p.
4. *Shannon C.E.* A Mathematical Theory of Communication // Bell System Technical Journal. (1948) T. 27. pp. 379-423, 623–656.
5. *Gurevich I. M.* The laws of informatics-the basis of structure and knowledge of complex systems. (In Russian). Torus Press. Moscow. 2007. 400 p.
6. *Gurevich I.M.* Structure of the Universe with the minimum information. (In Russian) Works of conference BAK-2007. Kazan. (2007), pp. 432-434.
7. *Gurevich I.M.* About restrictions on content of the information of the Universe. 58 th International Astronautical Congress – 2007 (IAC 2007),
8. *Gurevich I.M.* On information models in cosmology. Systems and Means of Informatics. Issue 17. IPI RAN. Moscow. 2007. pp. 164-183.
9. *Brillouin L.* Science and information theory. (In Russian) Fizmatgiz. Moscow (1960), 392 p.
10. *Valiev K.A. Kokin A.A.* Quantum computers: Hope and reality. (In Russian) Scientific and publishing center "Regular and chaotic dynamics". Moscow-Izhevsk (2004), 320 p.
11. *Kokkedee J.J.* The quark model. University of Nijmegen. The Netherlands. W.A.Benjamin,inc., New York – Amsterdam. 1969.
12. *Feynman R.P., Leighton R.B., Sands M.* The Feynman lectures on phisics. V.3. Addison-Westly publishing company, inc., Reading, Massacgusetts, Palo alto, London. 1965.
13. *Lloyd Seth.* Computational capacity of the Universe. arXiv:quant-ph/0110141 v1 24 Oct 2001.
14. *Vasiliev A.N.* Evolution of the Universe. The St.-Petersburg state University. (In Russian). http://www.astronet.ru/db/msg/1210286.